\documentclass[oldversion]{aa}
\usepackage{graphicx}
\usepackage{txfonts}
\usepackage{rotating}
\usepackage{lscape}
\newcommand{\gsim}{\;\lower.6ex\hbox{$\sim$}\kern-7.75pt\raise.65ex\hbox{$>$}\;}
\newcommand{\lsim}{\;\lower.6ex\hbox{$\sim$}\kern-7.75pt\raise.65ex\hbox{$<$}\;}

\begin{document}
\title{Intrinsic iron spread and a new metallicity scale for Globular
Clusters\thanks{Based on observations  collected at ESO telescopes under
programmes 072.D-507 and 073.D-0211}
 }

\author{
Eugenio Carretta\inst{1},
Angela Bragaglia\inst{1},
Raffaele Gratton\inst{2},
Valentina D'Orazi\inst{2},
\and
Sara Lucatello\inst{2,3}
}

\authorrunning{E. Carretta et al.}
\titlerunning{Iron spread and a new metallicity scale of globular clusters}

\offprints{E. Carretta, eugenio.carretta@oabo.inaf.it}

\institute{
INAF-Osservatorio Astronomico di Bologna, Via Ranzani 1, I-40127
 Bologna, Italy
\and
INAF-Osservatorio Astronomico di Padova, Vicolo dell'Osservatorio 5, I-35122
 Padova, Italy
\and
Excellence Cluster Universe, Technische Universität München, 
 Boltzmannstr. 2, D-85748, Garching, Germany 
  }

\date{Received 28 July 2009 / Accepted 1 October 2009}

  \abstract{
  We have collected spectra of about 2000 red giant branch (RGB) stars in 19
  Galactic globular clusters (GC) using FLAMES@VLT (about 100 star with GIRAFFE
  and about 10 with UVES, respectively, in each GC).  These observations provide
  an unprecedented, precise, and homogeneous  data-set of Fe abundances in GCs.
  We use it to study the cosmic scatter of iron and find that, as far as Fe is
  concerned, most GCs can still be considered mono-metallic, since the upper limit to
  the scatter in iron is less than 0.05 dex, meaning that the degree of
  homogeneity is better than 12\%.  
  The scatter in Fe we find seems to have a dependence on luminosity, possibly
  due to the well-known inadequacies of stellar atmospheres for upper-RGB 
  stars and/or to intrinsic variability.
  It also seems to be correlated with cluster properties, like
  the mass, indicating a larger scatter in more massive GCs which is likely a
  (small) true intrinsic scatter.
  The 19 GCs, covering the metallicity range of the bulk of Galactic GCs, 
  define an accurate and updated  metallicity scale. We provide transformation 
  equations for a few existing scales. We also provide new values of [Fe/H], on our
  scale, for all GCs in the Harris' catalogue. 
}

\maketitle

\section{Introduction}
Nowadays, the chemical composition of globular cluster (GC) stars cannot be
regarded anymore as strictly homogeneous, as implied from the classical paradigm
that these old stellar aggregates were the best approximation in nature of
simple stellar populations (see Gratton, Sneden and Carretta 2004 for an extensive
review and references). Recent investigations using large samples of stars
observed with multi-object spectrographs at large telescopes highlight that every
GC studied so far harbours at least two different stellar generations, distinct
in chemical composition and age (Carretta et al. 2009a,
2009b).

These different populations are found to differ in abundances of light elements 
(C, N, O,  Na, Mg, Al, Si,  F; Smith and Martell 2003; Smith et
al. 2005; Carretta et al.
2009a,b; Yong et al. 2008a,b; Melendez and Cohen 2009, to quote a few studies; see
also  Gratton et al. 2004 and references therein) involved in proton-capture
reactions of
H-burning at high temperature (Denisenkov and Denisenkova 1989; Langer et al.
1993). Star to star abundance variations in these elements are expected
to also come with differences
in the main outcome of the H burning, the He content (Gratton et al.
2009;  Bragaglia et al. in preparation; Prantzos and Charbonnel 2006; Ventura et al.
2001). Apart from a few alterations presently well understood in term of an
extra-mixing episode after the red giant branch (RGB) bump (Charbonnel 1994,
1995; Charbonnel and Zahn 2007;  Eggleton et al. 2007) the abundance variations
are inherited by currently observed, long  lived GC stars from a previous
stellar component/generation: both spectroscopic (Gratton et al. 2001; Ramirez
and Cohen 2002; Carretta et al. 2004; Piotto et al. 2005) and photometric  (e.g.
Bedin et al. 2004) observations convincingly showed that the observed pattern of
chemical composition is present also among unevolved stars on the subgiant
branch and the main sequence.

On the other hand, apart from a few notable exceptions\footnote{The GC $\omega$
Cen (widely considered to be the remnant of a dwarf galaxy) shows evidences of
several bursts of star formations, with corresponding peaks in the metal
abundance (see Gratton et al. 2004 for a comprehensive discussion) and a spread
in metallicity seems to be confirmed in M~22  (Marino et al. 2009) as well as
in M~54 (Bellazzini et al. 2008 and Carretta et al., in preparation).} GCs
are still found mono-metallic objects, as far as abundances of heavier elements
are concerned (see Gratton et al. 2004 for a recent review on this subject).
Their heavy (Z$>$13) elements metallicity, usually represented by the ratio 
[Fe/H]\footnote{We adopt the usual spectroscopic notation, $i.e.$ 
[X]= log(X)$_{\rm star} -$ log(X)$_\odot$ for any abundance quantity X, and 
log $\epsilon$(X) = log (N$_{\rm X}$/N$_{\rm H}$) + 12.0 for absolute number
density abundances.}, is found to be extremely homogeneous from star to star 
in each cluster.

The sites of production of heavy elements (in particular $\alpha-$capture elements, Fe-group
elements) are stars with large and intermediate initial masses, exploding as core-collapse 
or thermonuclear supernovae (see Wheeler, Sneden and Truran 1989). By studying the level and the dispersion of their yields
inherited by stars in GCs we have the possibility of investigating the past history of
the precursors from whom the present-day globular clusters formed (see Carretta et al. 
2009c). 

The recently completed analysis of high resolution spectra of almost 2,000 RGB
stars in  19 Galactic GCs (Carretta et al. 2009a,b and references
therein) provides an unprecedented sample of  stars
analysed in a fully homogeneous way. By exploiting these data we can examine in
detail the issue of the cosmic scatter in the metallicity (hereafter the
abundance of Fe-peak elements) of GCs and
hopefully provide clues to the early evolution of their progenitors.

Moreover, abundances of iron from high resolution spectra are traditionally the
calibrating points for metallicity scales based on photometric and/or
low-resolution spectroscopic indices that are sensitive to metal abundances (see
the excellent discussion and historical review on this issue by Kraft and Ivans
2003).  Thus, our sample is perfectly suited to provide robust (for statistics)
and homogeneous  (for technique of analysis) "pillars" onto which anchor several existing
metallicity scales. 

The paper is organized as follows.
We give in section 2 a brief description of the data set and of the analysis.
We discuss the intrinsic (cosmic) spread of iron in section 3, where we also
present some correlations of the spread with GC properties, like the absolute
total luminosity or the $\alpha-$elements abundance. We discuss the dependence
of this spread on the stars luminosity in section 4 and present a recalibration
of four metallicity scales to our new one in section 5. A summary is given in
section 6 and we give metallicity values on our new scale for all GCs in the
Harris (1996) catalogue in the Appendix.
 
\section{The dataset and analysis}

Briefly, in our survey we collected GIRAFFE spectra of intermediate resolution
($R\simeq 20000$) for about 100 stars and UVES spectra ($R\simeq 40000$) for
about 10 stars, on average, in each of our 19 programme GCs. The GIRAFFE
gratings were chosen to include the Na {\sc i} lines at 568.2-568.8\,nm (grating
HR11) and the forbidden [O {\sc i}] lines at 630.0-636.3\,nm (HR13).  The number
of clean Fe {\sc i} lines typically falling in the corresponding spectral
ranges, and used in our analysis, varied from about 10 to a maximum of about 40,
depending on the S/N, the metallicity and whether the stars were observed with
HR11 only, HR13 only, or both. The UVES spectra covered the 480-680\,nm region and
plenty of Fe lines were measured even in the most metal-poor clusters. 

Data analysis has already been described in detail elsewhere (Carretta et al. 2006,
2007a,b,c; Gratton et al. 2006, 2007; Carretta et al. 2009a,b) and will not be 
repeated here. Only a few procedures will be briefly summarised here.
Atmospheric parameters were homogeneously derived for all stars using visual
and near-IR photometry and the relations given by Alonso et al. (1999, 2001).
Equivalent widths ($EW$s) were measured by an updated semi-automatic routine,
in the
package ROSA (Gratton 1988), using an homogeneous method for the continuum
positioning (described in Bragaglia et al.  2001). The line list is
adopted from Gratton et al. (2003), where the atomic parameters and  solar
reference abundances are fully discussed. This list was used for all stars in
our program clusters, and we believe that this  represents a remarkable strength
of this project, resulting in  a huge database of stellar abundances derived in
the most homogeneous way for a significant part ($\sim 12\%)$ of the Galactic 
GC population, spanning the range in [Fe/H] from -2.5 to -0.4 dex.

As discussed in the papers quoted above, all $EW$s measured on the GIRAFFE
spectra were  shifted to a system defined by those derived from high
resolution UVES spectra, using stars and lines observed with both instruments.

\begin{table*}
\caption{Average abundances of [Fe/H] {\sc i} and observed rms scatters in our sample}
\small
\begin{tabular}{llrllrllll}
\hline
GC       & alt. & nr.   &    [Fe/H] {\sc i}       & $rms$& nr.&    [Fe/H] {\sc i}        &$rms$&GIRAFFE   &UVES       \\
         &name  &stars  & $\pm$stat.err$\pm$syst.&      &stars& $\pm$stat.err$\pm$syst.&     &          &           \\
         &      &GIRAFFE&       (dex)            &      &UVES&      (dex)              &     &          &           \\ 
\hline
NGC  104 &47~Tuc& 147 & $-0.743\pm0.003\pm0.026$ & 0.032& 11 & $-0.768\pm0.016\pm0.031$&0.054&Paper VII &Paper VIII \\
NGC  288 &      & 110 & $-1.219\pm0.004\pm0.070$ & 0.042& 10 & $-1.305\pm0.017\pm0.071$&0.054&Paper VII &Paper VIII \\
NGC 1904 &M~79  &  58 & $-1.544\pm0.005\pm0.069$ & 0.036& 10 & $-1.579\pm0.011\pm0.069$&0.033&Paper VII &Paper VIII \\
NGC 2808 &      & 123 & $-1.104\pm0.006\pm0.046$ & 0.065& 12 & $-1.151\pm0.022\pm0.050$&0.075&Paper I   &Paper VIII \\
NGC 3201 &      & 149 & $-1.495\pm0.004\pm0.073$ & 0.049& 13 & $-1.512\pm0.018\pm0.075$&0.065&Paper VII &Paper VIII \\
NGC 4590 &M~68  & 122 & $-2.227\pm0.006\pm0.068$ & 0.071& 13 & $-2.265\pm0.013\pm0.070$&0.047&Paper VII &Paper VIII \\
NGC 5904 &M~5   & 136 & $-1.346\pm0.002\pm0.062$ & 0.023& 14 & $-1.340\pm0.014\pm0.064$&0.052&Paper VII &Paper VIII \\
NGC 6121 &M~4   & 103 & $-1.200\pm0.002\pm0.053$ & 0.025& 14 & $-1.168\pm0.012\pm0.054$&0.046&Paper VII &Paper VIII \\
NGC 6171 &M~107 &  33 & $-1.065\pm0.009\pm0.026$ & 0.044&  5 & $-1.033\pm0.029\pm0.038$&0.064&Paper VII &Paper VIII \\
NGC 6218 &M~12  &  79 & $-1.310\pm0.004\pm0.065$ & 0.033& 11 & $-1.330\pm0.013\pm0.066$&0.042&Paper IV  &Paper VIII \\
NGC 6254 &M~10  & 147 & $-1.556\pm0.004\pm0.074$ & 0.053& 14 & $-1.575\pm0.016\pm0.076$&0.059&Paper VII &Paper VIII \\
NGC 6388 &      &  36 & $-0.406\pm0.013\pm0.028$ & 0.078&  7 & $-0.441\pm0.014\pm0.025$&0.038&Paper VII &Paper VI   \\
NGC 6397 &      & 144 & $-1.993\pm0.003\pm0.060$ & 0.039& 13 & $-1.988\pm0.012\pm0.061$&0.044&Paper VII &Paper VIII \\
NGC 6441 &      &  25 & $-0.332\pm0.017\pm0.040$ & 0.087&  5 & $-0.430\pm0.026\pm0.050$&0.058&Paper V   &Paper III  \\
NGC 6752 &      & 137 & $-1.561\pm0.004\pm0.073$ & 0.041& 14 & $-1.555\pm0.014\pm0.074$&0.051&Paper II  &Paper VIII \\
NGC 6809 &M~55  & 156 & $-1.967\pm0.004\pm0.072$ & 0.044& 14 & $-1.934\pm0.017\pm0.074$&0.063&Paper VII &Paper VIII \\
NGC 6838 &M~71  &  39 & $-0.808\pm0.005\pm0.048$ & 0.034& 12 & $-0.832\pm0.018\pm0.051$&0.061&Paper VII &Paper VIII \\
NGC 7078 &M~15  &  84 & $-2.341\pm0.007\pm0.067$ & 0.061& 13 & $-2.320\pm0.016\pm0.069$&0.057&Paper VII &Paper VIII \\
NGC 7099 &M~30  &  64 & $-2.359\pm0.006\pm0.067$ & 0.046& 10 & $-2.344\pm0.015\pm0.069$&0.049&Paper VII &Paper VIII \\
\hline
\end{tabular}
\label{t:ferms}
\\
Columns 9 and 10 provide the references of the papers where the analysis of all GIRAFFE and
UVES data, respectively, is presented for each cluster.

\end{table*}

Taking into account the stars in common between UVES and GIRAFFE observations,
we have values of [Fe/H] for 1958 individual RGB stars in 19 GCs.
Average abundances of iron derived from both UVES and GIRAFFE
spectra are summarised  in Tab.~\ref{t:ferms}. Only abundances derived from
neutral Fe lines are reported (and will  be used in the following) because we
already showed in Paper VII and VIII that the differences in iron abundances as
obtained from neutral and singly ionised species are negligible in our 
analysis\footnote{Note that we usually avoided observations of stars close to
the RGB tip, that may be more problematic.}.
Thus, we will adopt the [Fe/H] {\sc i} ratio as representative of the
metallicity of each star, because it is based on a much larger number of lines, thus
providing a more robust estimate.

\begin{figure}
\centering
\includegraphics[bb=19 145 337 710, clip, scale=0.59]{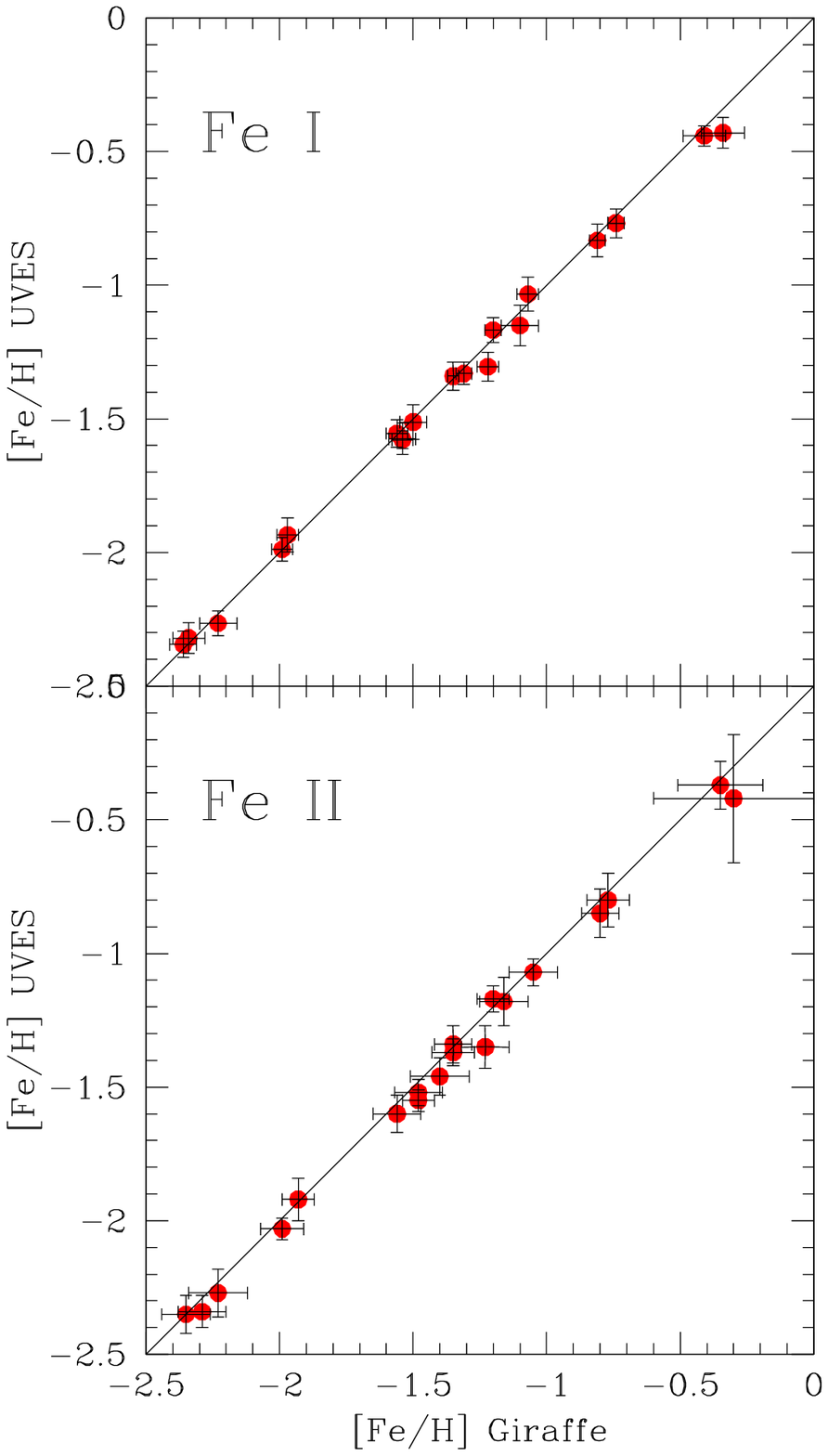}
\caption{Comparison for [Fe/H] {\sc i} ratios (upper panel) and for
[Fe/H] {\sc ii} ratios
(lower panel) obtained from GIRAFFE and UVES spectra in the 19 programme
clusters in our sample. Metal abundances from GIRAFFE are from Papers I,II,IV,V,
and VII. Metallicities from UVES are from Paper III, VI, and VIII.
Error bars are 1$\sigma$ $rms$ scatter and the solid line is the line of
equality.}
\label{f:FeIFeII}
\end{figure}

Once $EW$s measured from GIRAFFE spectra are corrected to the system defined by
UVES spectra, the abundance analysis provides very similar results for the
average metallicity in each cluster: the mean difference (in the sense UVES
minus GIRAFFE) is $-0.015\pm0.008$ dex with an $rms$ scatter of 0.037~dex from
19 GCs. The largest difference (0.098 dex) is found for NGC~6441, for which
the observations in service mode could not be completed as requested, so that
the available spectra have low S/N, due both to the large
distance modulus and incomplete observations.
Fig.~\ref{f:FeIFeII}, reproduced from Paper VIII, shows the quite good agreement
between  abundances of iron as obtained from both instruments; thus, although we
will use  [Fe/H] {\sc i} values from the high resolution UVES spectra, we are
confident that the new metallicity scale presented in the present paper 
rests ultimately on {\it very accurate and homogeneous abundances
for a sample of almost 2000 cluster stars}.

\section{Intrinsic spread of iron in Globular Clusters}

\begin{figure}
\centering
\includegraphics[scale=0.45]{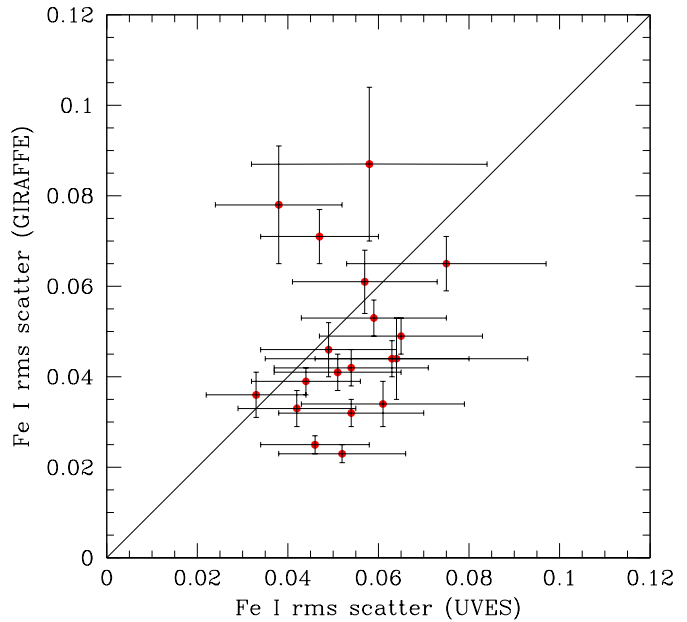}
\caption{Comparison of the intrinsic spreads in [Fe/H] for the 19 GCs 
in our sample, as derived from UVES and GIRAFFE spectra. Error bars
show the maximum and minimum spread allowed for each cluster by the statistical 
errors. The line indicating equality is also shown.}
\label{f:doss}
\end{figure}

We compare in Fig.~\ref{f:doss} the {\it bona fide} observed intrinsic spreads 
in Fe (defined as the $rms$ scatter of all stars in each cluster), as given
by GIRAFFE and UVES spectra (columns 5 and 8 in Tab.\ref{t:ferms}). The error
bars show the maximum and minimum spread allowed for each cluster taking into
account the statistical errors.

Looking at this figure, two features are immediately evident: first, on average,
the $rms$ scatter obtained from UVES spectra is larger than the one we derived
from the analysis of GIRAFFE spectra. At first sight, this result is just the
opposite one would expect, given the higher resolution and spectral coverage of
the UVES spectra. Possible explanations for this effect will be examined in the
next section.

Second, intrinsic scatters from GIRAFFE spectra show quite small associated
error bars, owing to the much larger statistics; if we focus on these values, the
first conclusion is that {\it the cosmic scatter in Fe in globular
clusters is very small}. On average, we found a value of
0.048 dex (with $\sigma=0.018$ dex) from 19 GCs; {\it i.e.} the iron abundance in each
cluster is homogeneous within 12\%.

Furthermore, we note that these are strictly {\it upper limits} to the actual
dispersion of [Fe/H] values in GCs. The true intrinsic values should be
estimated by subtracting (in quadrature) the scatter expected from errors in
the analysis, namely in the derivation of the atmospheric parameters and in the
measurement of $EW$s. 
These quantities were estimated with a thorough procedure amply described in
previous papers and are reported in the Appendix of Paper VII and in Papers I to
VI. Unfortunately, for the GIRAFFE spectra these estimates often exceed the
observed scatter, probably because of the overestimate of the effect of some error sources,
making difficult to derive the intrinsic scatter by deconvolving for this
uncertainty. For 8 out 19 GCs this deconvolution was possible and from this,
as well as from the analysis of UVES spectra, 
we found that a
reasonable estimate is that the observed values of the $rms$ scatter must be
further reduced by 1-2\%, to account for the analysis uncertainties.

To obtain a larger statistical sample we considered the observed scatters, noting however
that they are upper limits. Therefore, our first conclusion is that, as far as
iron is concerned,  at first approximation the GCs can still be considered
mono-metallic aggregates, apart from the few exceptions mentioned in the
Introduction.

\begin{table}
\centering
\small
\caption{Pearson's correlation coefficients, degrees of freedom and 
significance for relations with iron spread in our GC sample}
\begin{tabular}{llll}
Parameter                        & $r_p$ &d.o.f.&signif.  \\
\hline     		
$M_V$                            &$-$0.49& 17& 95-98\%    \\
R$_{\rm perig.}$                 &  +0.47& 15& 90-95\%    \\
log T$_{\rm eff}^{\rm max}$ (HB) &  +0.38& 17& 90\%       \\
x(MF)                            &  +0.60& 11& 95-98\%    \\
rel.Age                          &$-$0.49& 17& 95-98\%    \\
$\sigma$(h)                      &  +0.59& 16& 99\%       \\
P	                         &  +0.65& 17& $>$99\%    \\
I	                         &$-$0.67& 17& $>$99\%    \\
E	                         &  +0.52& 17& $\sim$98\% \\
$[$(Mg+Al+Si)/Fe$]$              &$-$0.48& 16& 95-98\%    \\
$[<$Si,Ca$>$/Fe]                 &$-$0.70& 17& $>$99\%    \\
$[$Ca/Fe$]$ 	                 &$-$0.78& 17& $>$99\%    \\
$[<\alpha]$/Fe]                  &$-$0.72& 17& $>$99\%    \\
$[$Mg/Fe$]_{\rm max}$            &$-$0.52& 17& $~$98\%    \\
$[$Mg/Fe$]_{\rm min}$            &$-$0.56& 17& 98-99\%    \\
$[$Si/Fe$]_{\rm min}$            &$-$0.44& 17& 90-95\%    \\
\hline 
\label{t:corrrmsg}
\end{tabular}
\begin{list}{}{}
\item[-] $M_V$ from the on-line version of the Harris (1996) catalogue
\item[-] peri-galactic distance R$_{\rm perig.}$ in kpc from Dinescu et al. (1999) 
or Casetti-Dinescu et al. (2007)
\item[-] maximum temperature along the HB log T$_{\rm eff}^{\rm max}$(HB) 
from Recio Blanco et al. (2006) or derived in Carretta et al. (2009c)
\item[-] exponent x of the power-law relation for global Mass Function from
Djorgovski et al. (1993) 
\item[-] relative age parameter from De Angeli et al. (2005) and Rosenberg et al.
(1999)
\item[-] velocity dispersion at the half-mass radius $\sigma$(h) from Gnedin et
al (2002)
\item[-] fractions of primordial P, intermediate I and extreme E components from
Carretta et al. (2009a; Paper VII)
\item[-] the other abundance ratios are derived from UVES spectra (Carretta et al.
2009b; Paper VIII); here [$\alpha$/Fe] is the average of [Mg/Fe]$_{\rm max}$, 
[Si/Fe]$_{\rm min}$ and [Ca/Fe]
\end{list}
\end{table}

However, the high degree of homogeneity in our analysis, coupled with  our very large
sample, allow us to highlight subtle features that might be lost in the noise
when collecting data from separate sources.
In particular, we found several statistically significant relations between the
intrinsic iron spread and global cluster parameters. In Table~\ref{t:corrrmsg}
we summarise the Pearson's linear correlation coefficients, the degrees of 
freedom and the statistical significance for a number of these relations.

\begin{figure}
\centering
\includegraphics[bb=19 144 342 715, clip, scale=0.52]{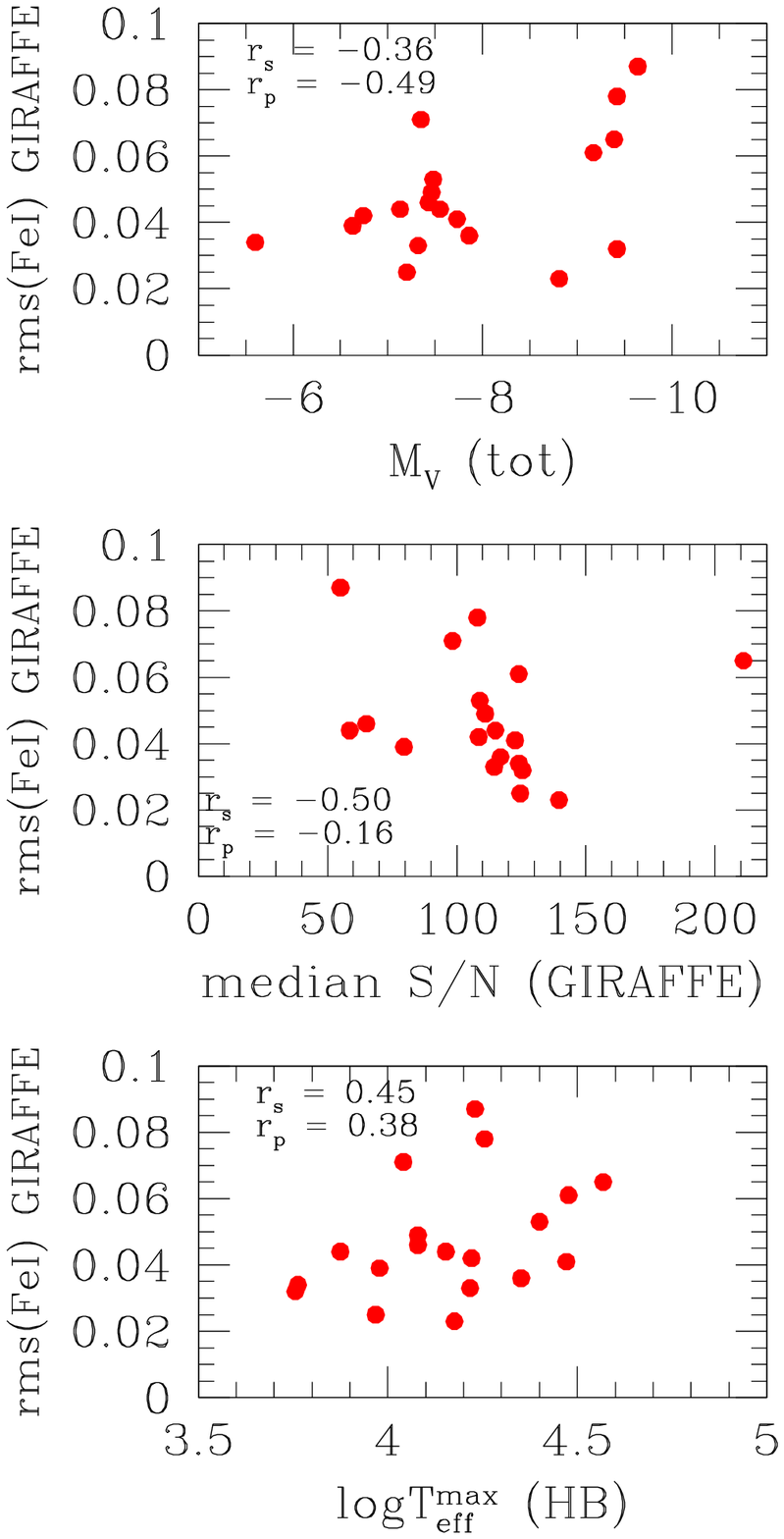}
\caption{Observed intrinsic dispersion in iron from GIRAFFE spectra as a
function of the visual total absolute magnitude of the clusters (upper panel),
of the median $S/N$ for HR13 spectra (middle panel),and of the maximum
temperature along the horizontal branch (lower panel). In  each panel of this
and next figures we list the Spearman and Pearson correlation coefficients.}
\label{f:rmsmass}
\end{figure}

Our main finding is that the observed dispersion in iron is correlated with the
cluster absolute magnitude $M_V$ (upper panel in Fig.~\ref{f:rmsmass}), a proxy for
the cluster present-day mass. More massive clusters show a larger spread in
[Fe/H]. Note that this correlation would extend to brighter cluster absolute
magnitudes if we include very massive clusters like M~54 and $\omega$ Cen.

The $rms$ seems preferentially larger in farther GCs, which could possibly
indicate that the trend may be driven by the data quality. However, this is
unlikely to be an observational artifact; as shown in the middle panel of
Fig.~\ref{f:rmsmass}, where we plot the median $S/N$ ratio  for the HR13
spectra, there is not a significant correlation between the quality of spectra
and the $rms$ scatter. Most clusters have a typical $S/N$ around 120, yet show
very different observed scatters. Moreover, our sample includes disk and inner
halo GCs, and in general more massive clusters are preferentially found at
larger distances (i.e. among the inner halo component) also in the global 
population of galactic GCs (Harris 1996).

Several theoretical models (Larson 1987, Suntzeff and Kraft 1996; D'Ercole et
al. 2008; Decressin et al. 2008) suggest that currently observed GCs are only
a small fraction in mass of the original parent structures where they
formed.
In Carretta et al. (2009c) we proposed a formation scenario for GCs
where a so-called $precursor$ raises the metallicity to a level which
corresponds to the primordial abundances currently observed in a cluster. 
In other words, in present-day GCs we are not seeing the contribution of massive SNe: the
level of iron and $\alpha-$elements is already homogeneously established 
in the gas by the explosion of core-collapse SNe in the precursor/proto-cluster.
Of course the idea of pre-enrichment in large fragments seeding the present galactic
GC population is not new (see Searle and Zinn 1978). In our
scenario, the second effect of SNe in the precursor is to trigger a burst of
star formation leading to the build up of the primordial population, a fraction
of which we can still observe in GCs, making up about 1/3 of the total cluster 
stars (see Carretta et al. 2009a).
However, the correlation we see between iron spread and cluster mass tells us
that clusters originating in more massive precursors are probably more capable to
retain inhomogeneities in the metallicity plateau established in the previous
phase of rapid enrichment.

This is just what is observed for the $rms$ scatter in [Fe/H] as a function of
$M_V$ and of other cluster parameters strictly related to the cluster mass: the
maximum temperature reached along the cluster HB (lower panel of
Fig.~\ref{f:rmsmass}; see Recio Blanco et al. 2006;
Carretta et al. 2009c), the slope of the cluster global mass function (see
Djorgovski et al. 2003), the velocity dispersion at half-mass radius
(Gnedin et al. 2002).

We note that a (small) part of the higher values for the spread in more massive
clusters may also be due to variations in He content, which are expected to be
larger in GCs of larger masses (Gratton et al. 2009). In this case, of course, 
the effect is due not only to the higher capability to retain 
ejecta but also to the enhanced He in some stars formed by gas heavily
polluted in H-processed material, which lowers the denominator in
the [Fe/H] ratio (see also Bragaglia et al. 2009). We can check the relevance
of this effect by computing the iron spread using only stars in the primordial
component P (see Carretta et al. 2009a), which are not expected to be formed from
matter polluted with additional He.

We found that the difference in the $rms$ of [Fe/H] (in the sense all stars
minus P stars) is on average $-0.002\pm0.003$  (with 
$\sigma=0.012$ for 19 GCs). For the two most massive clusters in the
sample (NGC~6388 and NGC~6441) this difference goes in opposite directions and
the largest differences are found for the three GCs with
lower quality spectra (NGC~6388 and NGC~6441 again, plus NGC~6752). We conclude
that the contribution to the spread in iron caused by variations in the He
content is probably very small.

\begin{figure}
\centering
\includegraphics[scale=0.42]{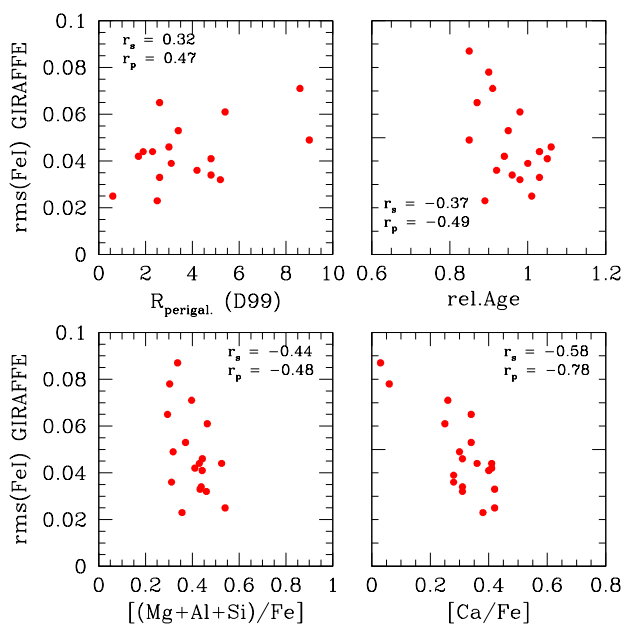}
\caption{Observed spread of [Fe/H] in our programme clusters as a function of 
their peri-galactic distance (upper left panel), age (upper right panel), 
total sum of Mg+Al+Si nuclei (lower left panel) and [Ca/Fe] ratio.}
\label{f:rmsposiz}
\end{figure}

A support for the proposed scenario comes from a second set of
(anti)correlations involving the observed spread of iron and the position
of GCs in the Galaxy (or related quantities; see
Tab.~\ref{t:corrrmsg} and Fig.~\ref{f:rmsposiz}). The dispersion
in [Fe/H] is larger in GCs spending, on average, more time at larger distances
from the Galactic centre (upper left panel). At the same time, the spread 
decreases for older clusters and for increasing abundances of the
$\alpha-$element\footnote{The [(Mg+Al+Si)/Fe] ratio represents the primordial
abundances of $\alpha-$elements in a cluster, mostly Mg and Si, as modified
by the following conversion of Mg into Al and, slightly, into Si in H-burning at
high temperature (see Carretta et al. 2009b).}.

In Carretta et al. (2009c) we emphasised that inner halo clusters
preferentially include
more massive clusters with respect to disc clusters; moreover,
the first are also younger, on average, and with a slightly smaller level of
$\alpha-$elements than that found in disc/bulge GCs. All these features concur
to explain the good correlation and the anti-correlations displayed in
Fig.~\ref{f:rmsposiz}.

\begin{figure}
\centering
\includegraphics[bb=36 144 320 714, clip, scale=0.52]{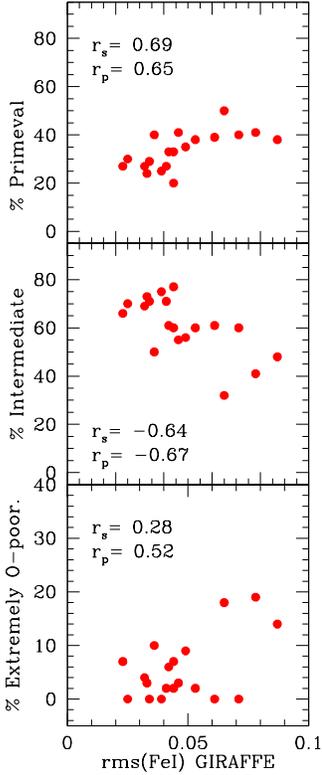}
\caption{Fraction of stars in the P (upper panel), I 
(middle panel), and E component (lower panel) of our programme clusters as
a function of the observed $rms$ scatter in [Fe/H].}
\label{f:rmspie}
\end{figure}

Finally, in Fig.~\ref{f:rmspie} we show the run of the $rms$ scatter in Fe
as a function of the fraction of first generation stars P, as well as
the second generation components of intermediate modified (I), and extreme 
modified (E) composition 
(see Carretta et al. 2009a for detailed definitions of these stellar 
populations).
The dispersion in Fe increases along with the fraction of P stars still
observed in GCs, again pointing toward a deeper potential well in more massive
clusters, more able to retain both SN ejecta and low mass P stars. On the
other hand, we know that the fraction of E stars is larger in more massive
clusters, and this explains also the correlation between the spread and the E
component. The complementarity between I and E fractions (Carretta et al. 2009c)
well accounts for the anti-correlation shown in the middle panel of
Fig.~\ref{f:rmspie}.

\section{Fe scatter as a function of the luminosity}

Coming back to the comparison of the $rms$ scatters in Tab.~\ref{t:ferms}, we
now look at the rather surprising fact that the $rms$ we derive for Fe 
abundances is higher from the high-resolution UVES spectra than from the
intermediate-resolution GIRAFFE ones. This effect is particularly clear for
clusters with metallicity $-1.4<$[Fe/H]$<-0.7$ dex in our programme
sample, which
have data of higher quality. More metal-rich clusters (e.g. NGC~6388 and
NGC~6441) are quite distant, bulge clusters, severely affected by field
contamination and required a series of long exposures which were not always
completed in the ESO observing service queue (in particular for NGC~6441).
Analysis of more
metal-poor clusters is somewhat hampered by the weakness of lines, increasing
the difficulty of $EW$ measurements.

To explain the above mentioned effect, a possibility is that stars observed with UVES have
intrinsically a larger scatter than those observed with GIRAFFE. This 
would happen if the intrinsic scatter is partly a function of the stellar 
luminosity, because we preferentially chose brighter stars as UVES targets
(although the majority of the objects - about 170 out of 214 stars - are in
common between the two data sets).

\begin{table}
\centering
\small
\caption{Average residuals with respect to the mean [Fe/H] for clusters with
metallicities $-1.4<$[Fe/H]$<-0.7$ dex}
\begin{tabular}{rrcl}
\hline
$M_K^0$ bin & stars& $<$[Fe/H]$_{star}$ -[Fe/H]$_{GC}>$ & $rms$ \\
\hline
                 & \multicolumn{3}{c}{GIRAFFE} \\
$+0.0 \div -1.0$ &  97 &$ +0.006\pm 0.004$& 0.035 \\
$-1.0 \div -2.0$ & 271 &$ +0.003\pm 0.002$& 0.033 \\
$-2.0 \div -3.0$ & 155 &$ -0.009\pm 0.003$& 0.038 \\
$-3.0 \div -4.0$ &  81 &$ +0.004\pm 0.006$& 0.065 \\
$-4.0 \div -5.0$ &  56 &$ -0.012\pm 0.006$& 0.043 \\
                 & \multicolumn{3}{c}{UVES} \\
$-1.5 \div -3.0$ & 16 &$+0.008\pm 0.014$& 0.056 \\
$-3.0 \div -4.0$ & 15 &$-0.004\pm 0.008$& 0.051 \\
$-4.0 \div -5.5$ & 30 &$-0.001\pm 0.010$& 0.058 \\
\hline 
\label{t:diffe}
\end{tabular}
\end{table}

To better understand this point we need to improve the estimate of the intrinsic
scatter and in order to clarify the issue we proceeded as follows. 
For each star observed with GIRAFFE or UVES (separately) we evaluated the
difference between the [Fe/H] value for the individual stars and the average
value [Fe/H] of the cluster. These residuals were then plotted as a function of
the absolute $K$ magnitude $M_K^0$ corrected for the apparent distance modulus
and reddening in each cluster. We chose the $K$ band because (i) all near-IR
magnitudes are from the 2MASS catalogue and they represent a very homogeneous
dataset and (ii) this filter is much less sensitive to errors in the reddening
estimate.

\begin{figure}
\centering
\includegraphics[scale=0.40]{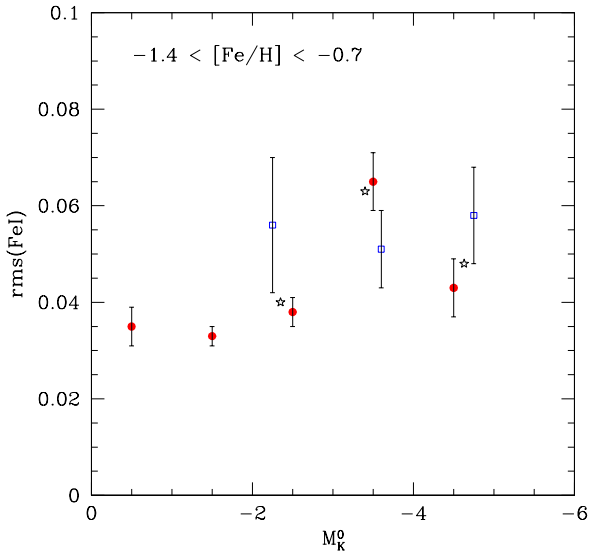}
\caption{Average differences between individual stars and the mean [Fe/H] value
in each cluster in the metallicity range $-1.4<$[Fe/H]$<-0.7$, as a function of
the near IR de-reddened absolute magnitude $M_K^0$. Filled circles: stars
observed with GIRAFFE; empty squares: stars observed with UVES. The open star
symbols are the weighted average of residuals from UVES and GIRAFFE in the
corresponding luminosity bin.}
\label{f:trend}
\end{figure}

Finally, all stars were binned in luminosity intervals and the average residual
and $rms$ scatter from the mean was computed for each bin. 
These averages are listed in Tab.~\ref{t:diffe} and plotted in 
Fig.~\ref{f:trend}. We restricted this exercise to the metallicity range [Fe/H]
from -1.4 to -0.7 dex, where we have better quality data.

We can see that the brighter stars (for which the $S/N$ ratio is higher) show a
larger scatter in the abundances of iron than the fainter stars. The difference
is quite significant, considering the errors associated to the estimate of the
scatter: in the bin $-2< M_K^0 <-1$, with 271 stars, we
obtain an $rms$ scatter of $0.033\pm 0.002$ from the GIRAFFE
observations, whereas for the brighter bin 
$-4< M_K^0 <-3$ (81 stars) we have $0.065\pm 0.004$. This difference is
statistically significant, at a level of more than 4$\sigma$.
The results from the UVES spectra are essentially consistent with those derived
from GIRAFFE (although with larger statistical errors owing to the much smaller
number of stars available) and reveal a clear trend of the $rms$ as a function
of the luminosity.

This trend can be explained with the reasonable assumption that the atmosphere
of each star (among
the more luminous objects) has a distinct peculiarity or intrinsic time
variability. Ita et al. (2002) found that very bright RGB stars in the LMC are
indeed variables (likely pulsating). However, the stars considered here are
considerably fainter, so that we are more inclined to attribute this variability
to the large sizes (and small numbers) of convective
cells with respect, e.g., to solar analogues\footnote{Note, however, that this
cannot be the full explanation. Studies of small amplitude variable red giants in LMC
and SMC from the huge OGLE-II database (Soszynski et al. 2004) reveal that
multi-periodic variability affects the majority of stars, with indications of
non-radial oscillations for a sizable fraction of objects.}.
As a consequence, there is an
uncertainty (of the order of about 0.04-0.05 dex) in the Fe abundances derived
from an instantaneous observation, since the convective elements of a red giant
are so large that only a small number of them occupy the surface of the star at
any time (Schwarzschild 1975; Gray et al. 2008). 
To account for this uncertainty we would need to assume that the temperature in
the line formation region differs from star to stars, with an $rms$ scatter of
about 50 K. 

While Schwarzschild (1975) estimated in about 400 the number of convective cells
in the atmosphere of a red giant, our sample is mostly formed by warmer stars.
We then assume that the number of convective cells is larger, in our case, say
about 1000. Of those, only half are obviously visible and not all of them with the
same weight, because of the limb darkening. Thus, we would get random fluctuations
of about a factor 20 smaller than those expected at the star surface between the
hotter regions (the top of the ascending columns) and the cooler regions
(descending columns), having temperature differences by about 1000 K.
Hence, we could expect that random fluctuations of the temperature
in the region of line formation in a giant are effectively of the order of 50 K,
even at similar effective temperatures.
Following the same line of thought, one could expect that the luminosity is not
perfectly constant, but should show fluctuations up to 0.01 mag (over
timescales of a few tens of days); the same would show up in fluctuations of
radial velocity (over similar timescales and amplitude of about 200-300 m/s).
These expectations seem to agree with observations of stars similar to those in
our sample (the threshold being $-2<M_V<-1.4$, Cummings 1999; Carney et al.
2008).

In summary, since the spread measured from brighter stars could include the
intrinsic noise due to the convection and pulsation (both in radial and
non-radial mode), the implication is that it is
better to use fainter stars in order to obtain the star-to-star scatter. From
these stars (practically restricted to those with GIRAFFE spectra) the typical
$rms$ in a cluster should be about 0.033 dex (see Tab.~\ref{t:diffe}), or less,
since this scatter value includes also the measurement errors.

Finally, we remark that the scatter seems to be larger for more metal-poor
clusters. We
speculate that this is in part due to possibly larger observational errors, but
in part it could be due to an increased transparency of the atmosphere, hence
pointing to a larger relevance of the 3d-structure.

\subsection{Alternative scenario: contamination from AGB}

An alternative explanation of the larger scatter observed at higher luminosities
(in particular in the magnitude bin $-4< M_K <-3$) could be
contamination from asymptotic giant branch (AGB) stars.
Were this the case, we would expect that the sequences are sufficiently well
separated (and the contamination negligible) at fainter magnitudes. On the other
hand, the temperature difference between RGB and AGB
is small, in the more luminous bin. However, the error would be larger where the two sequences are so close
that cannot be distinguished, but separated enough to give a wrong temperature
when the luminosity is used, as in the procedure adopted in our analysis (see
Carretta et al. 2009a for further details).

In this scenario, we would expect a distribution essentially Gaussian in form at
lower luminosities, where the contamination is negligible. At higher magnitudes,
the distribution should become asymmetric, with a tail toward lower abundances
given by the possible interlopers AGB stars for which the effective temperature
(hence the abundance) would be underestimated. This asymmetry should be more
pronounced in the bin $-4< M_K <-3$.

\begin{figure}
\centering
\includegraphics[scale=0.45]{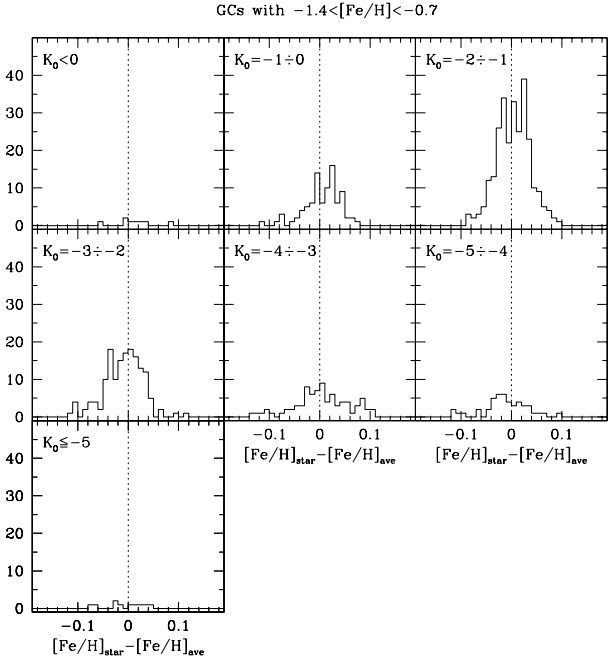}
\caption{Histograms of the residuals in Fe abundances in several magnitude bins
for clusters in our sample with metallicity $-1.4<$[Fe/H]$<-0.7$.}
\label{f:fek2}
\end{figure}

We checked this possible effect by computing the residuals in
Fe abundances derived from GIRAFFE spectra in different bins of magnitudes,
again restricting to the clusters in the metallicity range $-1.4<$[Fe/H]$<-0.7$
dex.
The histogram of the residuals in various magnitude bins are plotted in
Fig.~\ref{f:fek2} 

There seems to be no particular asymmetry
in the range $-4< M_K <-3$, where the distribution appears only slightly larger.
In conclusion, this test seems to support the idea that the larger spread is due
to the stellar atmospheres and to the variability provoked by the photospheric 
convection and stellar pulsation.

\section{A new homogeneous metallicity scale }

Having at hand an unprecedented large sample of GC red giants 
with homogeneously measured metal abundances, a logic step is to re-address the issue of 
an accurate metallicity scale for GCs. This is an indispensable
tool in several issues, like age derivation (both absolute and relative to other
clusters) of individual GCs as well as the
study of age-metallicity relation for cluster populations.

In the following we provide the relations to transform previous metallicity
scales against the one that can be defined on the basis of our UVES sample of
stars.

\paragraph{Recalibration of the Carretta \& Gratton (1997; CG97) scale -} Since
its appearance, this metallicity scale has been widely used due to the effort
made to use the most homogeneous approach. Briefly, the CG97 scale was derived
from the re-analysis of a large sample ($\sim 160$) of bright giants in 
24 GCs, whose abundances were obtained from $EW$s measured on high dispersion
(at the time, $R\sim30,000$) CCD spectra, adjusted to a common system. All
data were re-analysed using a common set of model atmospheres (the same we are
using in the present study, from Kurucz 1993), a common set of high quality
laboratory oscillator strengths $gf$s for Fe~I lines, and, whenever possible,
the same colour-temperature scale (Cohen, Frogel and Persson 1978), based on the
visual-near infrared colour $V-K$ available at the time. An improvement on
that seminal work is that in the present analysis $all$ the
$EW$s are measured with the same procedure, from higher resolution
($R\sim43,000$) spectra all acquired with the same instrument (UVES). Moreover,
atmospheric parameters T$_{\rm eff}$ and $\log g$ are obtained for all stars
using the same calibrations (on the Alonso et al. 1999 scale, based on the
infrared flux-method) and the same accurate $K$ magnitudes from the recent 2MASS
database. 

\begin{figure}
\centering
\includegraphics[scale=0.45]{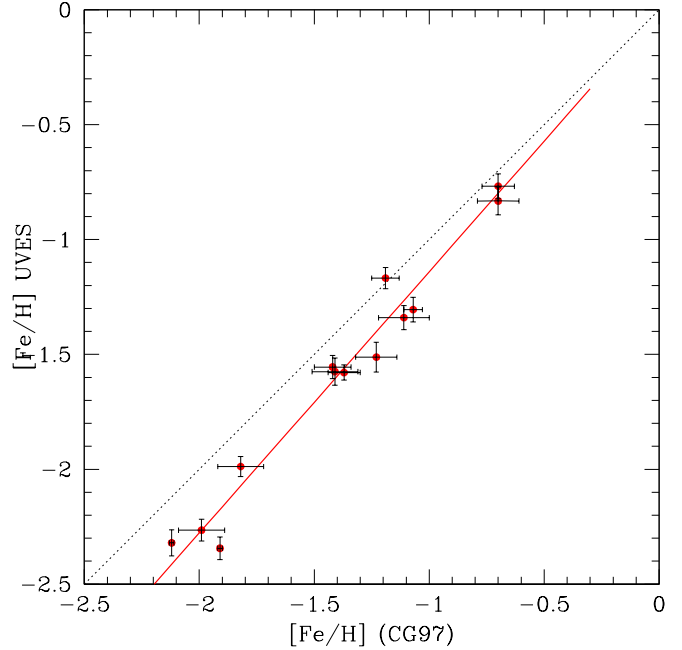}
\caption{Calibration of the CG97 metallicity scale vs average metal abundances
derived by us from UVES spectra for 13 clusters in common. The dotted line is
the equality line, the solid line is the linear regression obtained by
least-squares fit to the points. Error bars are $rms$ scatters of the mean.}
\label{f:cg97u}
\end{figure}

We used 13 clusters in common between CG97 and the present analysis to provide 
a new calibration of the metallicity scale. The regression line is
obtained by least-squares fit, by averaging
values derived exchanging the independent and dependent variables, and is
$${\rm [Fe/H]_{UVES}} = 1.137(\pm 0.060){\rm [Fe/H]_{CG97}} -0.003$$
with $\sigma=0.094$ dex and correlation coefficient $r=0.98$ from 13 clusters.
The comparison is shown in Fig~\ref{f:cg97u}.

Our new [Fe/H] values are systematically lower (by $0.19$ dex on average, with
$\sigma=0.11$ dex) than those by CG97, and the difference increases with
decreasing [Fe/H]. We must note, however, that the average values for the three
most metal-poor clusters (NGC~4590, NGC~7078, and NGC~7099) are based in CG97
only on a few stars (from 2 to 4) in each cluster, compared to more than 10
stars per cluster in our new analysis. The above relation shows that there is
almost only a multiplicative constant between the two scales and provides the
transformation required to bring the CG97 scale onto the new system.

Part of the difference is certainly due to the adopted temperature scale. Since
we observed different stars (farther away from the RGB tip in the present
study) and different colours (2MASS rather than on the CIT system, Frogel et al.
1983) a very
precise comparison is not possible. Anyway, the Alonso et al. scale is cooler:
we found differences between 57 and 97 K, for the same $V-K$ colours and
different metallicities, with the larger difference found for more metal-poor
GCs. Using the new Alonso et al. scale we would have had lower abundances also
on the CG97 scale, by about 0.09-0.10 dex for the more metal-poor GCs and by
only 0.02-0.03 dex for the more metal-rich GCs, which accounts for most of the
difference we found between the two scales.
The metallicity found here for M~30 can be explained by the different
temperatures: a star of magnitude similar to the two examined in CG97 has a
temperature lower by about 140 K in our new analysis. This difference, larger
than the average one for metal-poor GCs (about 100 K), justifies the observed
offset. 

However, even including this correction we would still have an offset of about
0.08-0.12 dex. Part of this residual difference is due to the $gf$ values: those
currently adopted by us (see Gratton et al. 2003) are larger on average by
$0.026\pm 0.005$ dex. The remaining difference of 0.05-0.09 dex are probably
due to the $EW$s, measured on higher quality and higher resolution spectra 
and in a more homogeneous way in the present project.

Finally, we note that most of the scatter around the best fit is given (with
opposite sign) by two clusters, NGC~6121 (M~4) and NGC~7099 (M~30).
For the first case, the difference is essentially due to the
anomalous value of the ratio $A_V/E(B-V)$, which was taken into account
for this cluster in the present
programme. The difference between the classical value 3.1 (CG97) and 4.0 (used
in this work) implies a difference of 0.28 mag in the correction of $V-K$, which
in turn translates in a difference of about 200 K in the effective temperatures,
just enough to shift M~4 on the best fit regression given by the other GCs.

Concerning NGC~7099 (M~30), the original data adopted by CG97 were from Minniti
et al. (1993) who derived the effective temperatures from the spectra; these
temperature were adopted also in CG97, due to the lack of $V-K$ colours for the
two stars analysed. 

\paragraph{Recalibration of the Zinn \& West (1984; ZW) scale -} The other most
used metallicity scale for globular clusters is the one defined by Zinn and 
West (1984), based on a variety of integrated-light photometric and
spectroscopic indices calibrated from the few echelle photographic spectra
existing at the time. Its popularity was due to the ranking this scale provided
for the first time for a large number of clusters; in fact, all 19 GCs of our
project have a corresponding [Fe/H] entry in the ZW scale.

\begin{figure}
\centering
\includegraphics[scale=0.45]{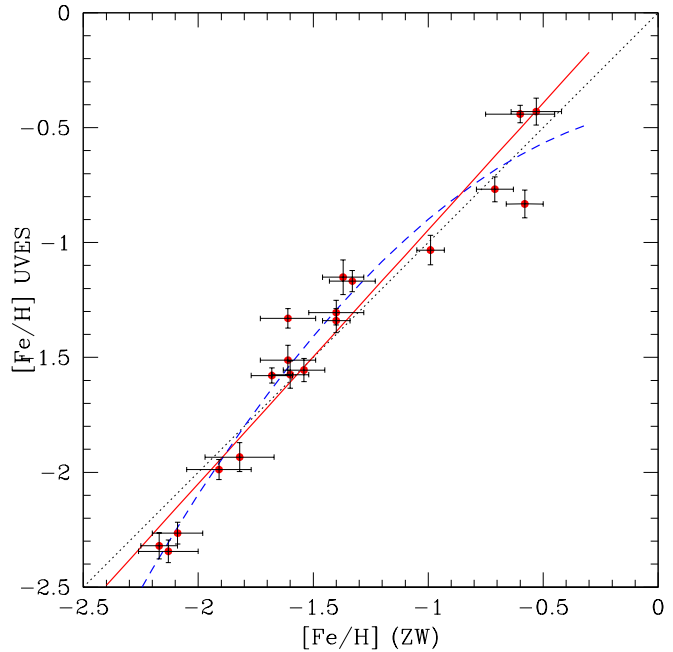}
\caption{Comparison of the ZW metallicity scale vs average [Fe/H] values
obtained by us from UVES spectra for all 19 clusters in our programme sample.
The dotted line is
the line of equality, the solid (red) line is the linear regression obtained by
least-squares fit to the points and the dashed (blue) line is a second-order
polynomial relation to the data. Error bars are $rms$ scatters of the mean.}
\label{f:zwu}
\end{figure}

The comparison between our average metallicities from UVES spectra and the ZW 
average [Fe/H] values is shown in Fig.~\ref{f:zwu}. A least-squares linear fit
(again obtained by exchanging the independent and dependent variables and averaging the
results) gives the following relation to transform ZW values to our new scale:
$${\rm [Fe/H]_{UVES}} = 1.105(\pm 0.061){\rm [Fe/H]_{ZW}} +0.160$$
with $\sigma=0.143$ dex and correlation coefficient $r=0.97$ from 19 clusters.

On average, the two scales formally differ by only 0.01 dex over the sampled
metallicity range, but the scatter of the points is now clearly larger than in
the case of the CG97 scale. By looking at Fig.~\ref{f:zwu} a better
transformation is obtained when a second-order polynomial is used:
\begin{eqnarray}
{\rm [Fe/H]_{UVES}} & = - 0.413 (\pm 0.027) + 0.130\,(\pm 0.289){\rm [Fe/H]_{ZW}} 
\nonumber \\
    & -0.356\,(\pm 0.108){\rm [Fe/H]_{ZW}}^2 \nonumber    
\end{eqnarray}
with $\sigma=0.119$ dex and correlation coefficient $r=0.98$ from 19 clusters.
We made an analysis of variance and tested with an F-test the statistical
significance of the regression with multiple components and the
significance of the coefficient of the highest degree. Both resulted highly
significant, and the resulting scatter in the transformation is quite decreased
by using the second order equation above.

We note however that $metallicities$ on the ZW scale are obtained by averaging
the results from a heterogeneous series of $indices$, calibrated on the best
metal abundances from high dispersion spectroscopy available at the time (Cohen
1983; Frogel et al. 1983). A better approach is to directly calibrate the
original and homogeneous $Q_{39}$ index by Zinn and West (1984), available for
a large sample of GCs, on the new metallicities presented here and use it in addition
to other homogeneous estimate of [Fe/H] to obtain an accurate ranking in metal
abundance for Galactic GCs. This is done in the Appendix.

\begin{figure}
\centering
\includegraphics[scale=0.45]{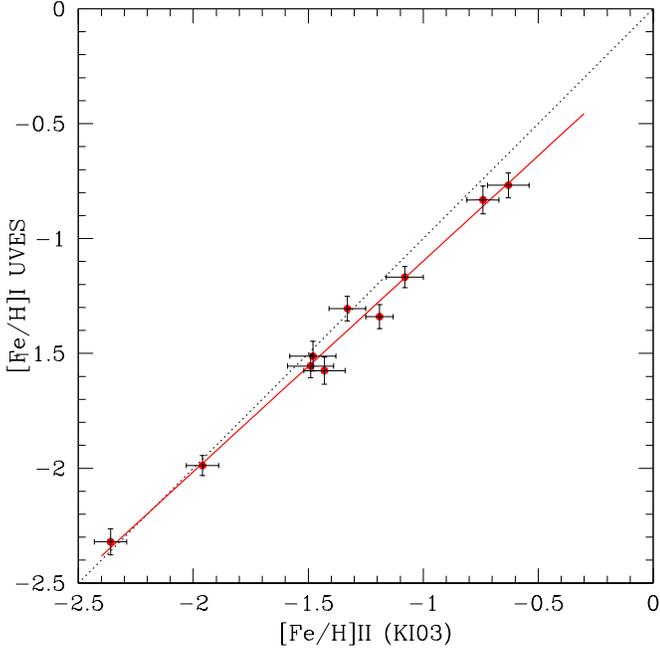}
\caption{Comparison of the Kraft and Ivans (2003) metallicity scale based on
Fe~II abundances vs  average [Fe/H] values obtained by us from Fe~I lines
measured on UVES spectra for 10 clusters in  common. The dotted line is the 
identity line, the solid (red) line is the linear regression obtained by
least-squares fit to the points. Error bars are $rms$ scatters of the mean.}
\label{f:ki03u}
\end{figure}

\paragraph{Recalibration of the Kraft \& Ivans (2003; KI03) scale -} In
Fig.~\ref{f:ki03u} we show the comparison of our [Fe/H] values 
from UVES
with those of the metallicity scale by Kraft and Ivans (2003), not much used
in spite of being based on
the homogeneous re-analysis of $EW$s of Fe~II lines from high resolution 
spectra. We adopted their values from the Kurucz model with convective
overshooting turned on (their Table 4, column 6) for consistency with the models
used in the present work.

The relation to transform the KI03 values on our new scale is:
$${\rm [Fe/H]_{UVES}} = 0.917(\pm 0.038){\rm [Fe/H]_{KI}} -0.181$$
with a scatter of only $\sigma=0.056$ dex and correlation coefficient 
$r=0.99$ from 10 clusters. Surprisingly, the agreement we found is quite good,
on average the two scale differ only by 0.07 dex (our averages being lower) but
this difference is scarcely significant (the scatter about the mean being 0.07
dex as well). The difference increases with increasing metallicity, and is on
average null at low [Fe/H] values.

This finding is rather strange because the basic reason for which Kraft and
Ivans used Fe~II is that it is generally assumed that currently
available one-dimensional LTE model atmospheres give a correct representation of
abundances from singly ionized Fe lines, but not of the abundances from Fe~I
lines, more affected by departures from the LTE assumption. Hence, they
concluded that the safest approach was to rely on Fe~II lines alone. 
However,
Fig.~\ref{f:ki03u} seems to question these conclusion for two related reasons.
First, because of the good
agreement we found on average, being our abundances derived from a neutral specie
and theirs from a singly ionized specie;
second, because departures from LTE
are expected to be larger where the stellar atmospheres become more transparent
and less line-blanketed, i.e. the effects increase toward lower metallicities,
whereas just the opposite is shown by the comparison.
At the moment, we have no clear explanation for this occurrence; we only note
that our samples are composed by stars not so close to the RGB tip, where NLTE
effects might be more relevant.

\paragraph{Recalibration of the near-infrared Ca~{\sc ii} scale (Rutledege et
al 1997; R97) -} Finally, we provide
here a new calibration of the metallicity scale based on the strength of the 
near-infrared Ca~{\sc ii}
triplet, used in particular to estimate the metallicity of stars in distant
systems, such as the nearby
dwarf spheroidals, because the triplet is measurable even at large distances.

\begin{figure}
\centering
\includegraphics[scale=0.45]{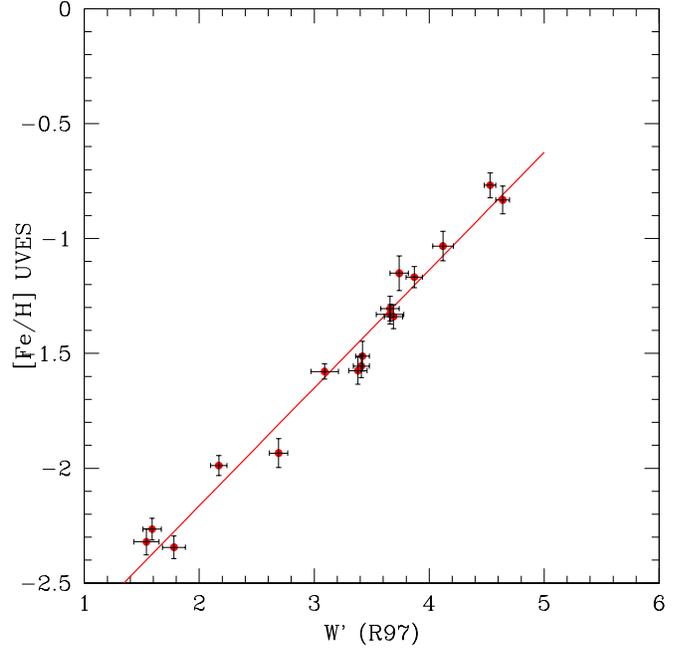}
\caption{Comparison of our [Fe/H] average values with the
reduced strength of the near-infrared Ca~II triplet $W'$ from Rutledge et 
al. (1997) for 17 clusters in common.
The superimposed solid (red) line is the linear regression obtained by
least-squares fit to the points. Error bars are $rms$ scatters of the mean.}
\label{f:r97u}
\end{figure}

In Fig.~\ref{f:r97u} we show the comparison between our UVES
metallicities and the values of the reduced
strength W'(Ca~{\sc ii}) of the near-infrared Ca triplet taken from Rutledge et al.
(1997; R97). The two variables are linearly correlated over the whole range
defined by the clusters analysed in the present paper 
($-2.4\lsim$ [Fe/H] $\lsim -0.8$; unfortunately, values of $W'$ are not given
 for NGC~6388 or NGC~6441. The transformation of the reduced
strength W' to the UVES metallicities 
$${\rm [Fe/H]_{UVES}} = 0.513(\pm 0.022) W'_{R97} -3.189$$
(with a correlation coefficient $r=0.99$ and a $rms$ scatter of only
$\sigma=0.086$ dex from 17 clusters) is therefore strictly only applicable up to
the metallicity of NGC~104 (47~Tuc), [Fe/H]=-0.77 dex. 
However, from the abundance analysis of two very metal rich bulge clusters, 
Carretta et al. (2001) already noted 
that the index $W'$ might deviate from linearity at high metallicity, whereas
in their work Rutledge et al. (1997) claimed that a linear relation existed
between W' and [Fe/H]$_{CG97}$, simply because results for metal-rich clusters
were not available at the time.
With the present sample we cannot extend the calibration further,
but in the Appendix we add two near solar metallicity bulge clusters 
to obtain a more extended calibration of this index, to be inserted in a final
compilation of GC metallicities.

\section{Summary}
Within our survey of 19 Galactic GCs with FLAMES (UVES and GIRAFFE), we have
measured the iron abundance of about 2000 RGB stars in a accurate and homogeneous
way (determination of atmospheric parameters, measure of $EW$s, spectroscopic
analysis, etc). In the present paper we  exploit this huge, homogeneous
data set, covering (almost) the entire range of GC metallicity in our Galaxy and
a summary of our findings follows:
\begin{itemize}
\item [i)] we obtain a measure of the intrinsic spread of iron in GCs from the
$rms$ scatter of our metallicities. We confirm that the cosmic scatter in Fe of
most GCs is very small: the upper limit to this scatter is less than 0.05 dex, i.e.,
the iron abundance is homogeneous within 12\% in each GC, on average;
\item [ii)] the $rms$ scatter we find for the UVES sample within each GC is
larger than the one derived for the GIRAFFE sample. This can
be ascribed to the different magnitude ranges observed. The UVES
stars are usually brighter, nearer the RGB tip, and may suffer from larger
inadequacies in the treatment  of atmospheres and intrinsic stellar variability;
\item[iii)] this is confirmed by a increase in scatter for brighter stars also
in the GIRAFFE samples, that is not explained by AGB contamination, and that is
larger for metal-poor GCs;
\item[iv)] there are interesting correlations between the $rms$ scatter in Fe
and several GC parameters, like $M_V$, $T_{eff}^{max}$(HB), $\alpha$-element 
abundance, that seem to indicate a better capability of more massive clusters
to self-enrich;
\item [v)] we recalibrate several metallicity scales of GCs using our 19
precise, homogeneous values; we have done it for the GC97, ZW, KI03, and R97
ones, giving transformation equations;
\item [vi)] finally, in the Appendix we give [Fe/H] on our new scale for all
clusters in the Harris (1996) compilation.
\end{itemize}

\begin{acknowledgements}
We wish to thank the ESO Service Mode personnel for their efforts. 
We also thank the referee for his/her useful comments. This
publication makes use of data products from the Two Micron All Sky Survey,
which is a joint project of the University of Massachusetts and the Infrared
Processing and Analysis Center/California Institute of Technology, funded by
the National Aeronautics and Space Administration and the National Science
Foundation. This work was partially funded by the Italian MIUR
under PRIN 2003029437. We also acknowledge partial support from the grant
INAF 2005 ``Experimental nucleosynthesis in clean environments". 
SL is grateful to the DFG cluster of excellence ``Origin and Structure of
the Universe" for support. 
\end{acknowledgements}

\begin{appendix}
\section{Cluster metallicities on a new homogeneous UVES scale}
In this Section, we provide an updated compilation of cluster metallicity
for all the Galactic GCs included in the Harris (1996) catalogue anchored
to modern measurements of stellar abundances
from high dispersion spectroscopy. 
We use as reference ``pillars" the accurate and homogeneous [Fe/H] values
derived in our project from high resolution UVES spectra for 214 giants stars in
19 GCs. The very good agreement between these abundances and those obtained for
stars observed also with GIRAFFE means that our scale is supported by the
observed metallicities of almost 2000 red giants, all analysed in the same
way.

In our approach we selected two sets of data to be included in our new 
calibration: 
\begin{itemize}
\item[(i)] recent and homogeneous metallicities from high resolution
spectroscopy of stars in a
limited number of GCs (CG97, 24 clusters; Kraft and Ivans 2003, 16 clusters)
\item[(II)] metallicity dependent indices homogeneously derived for large
samples of GCs ($Q_{39}$ from Zinn and West 1984, 60 clusters; W' from Rutledge
et al. 1997, 69 clusters).
\end{itemize}

Our metallicities are valid only over the metallicity range
sampled by our programme clusters. In the very low metallicity end this is not an
issue, since we analysed some of the most metal-poor clusters (NGC~7078,
NGC~7099, NGC~4590, and NGC~6397). However, problems might arise at the high
metallicity extreme, near solar [Fe/H] values. Since the most metal-rich
clusters are NGC~6388 and NGC~6441 ([Fe/H] about -0.4 dex),
metallicities beyond this limit would be derived in extrapolation, and this is a
risky procedure.

To alleviate this problem, we add to our 19 GCs two very metal-rich clusters,
whose metal abundances were newly analysed (or revised) in Carretta et al.
(2001): NGC~6528 ([Fe/H]$=+0.07$ dex, $\sigma=0.1$ dex) and NGC~6553
([Fe/H]$=-0.06$ dex, $\sigma=0.15$ dex). For these two clusters the
analysis is of course not strictly homogeneous with the other 19 GCs, however the $gf$
scale is very similar and their addition to the sample is unavoidable to
obtain a
reasonable calibration of the metal rich end of the GC scale, as
we will see below.

\begin{figure}
\centering
\includegraphics[scale=0.45]{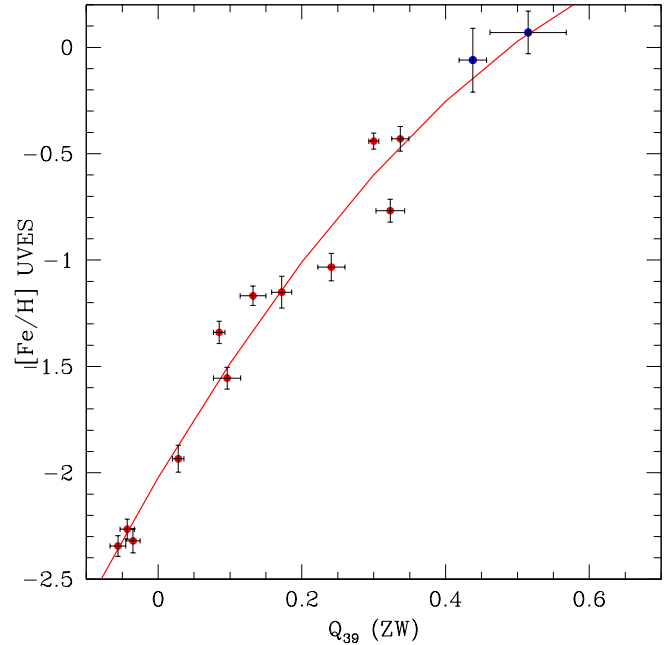}
\caption{Calibration of the $Q_{39}$ index from Zinn and West (1984) using
our [Fe/H] values from high resolution spectroscopy. Filled red circles are our
clusters with UVES spectra; blue filled squares are the two metal-rich bulge
clusters whose metallicities were analysed (or revised) in Carretta et al.
(2001), based on Keck HIRES spectra.}
\label{f:q39}
\end{figure}

In Fig.~\ref{f:q39} we show the calibration of the $Q_{39}$ index from Zinn and West (1984) using
the 12 GCs with UVES metallicities in common with that work and the two added clusters NGC~6528
and NGC~6553. From this figure it is clearly evident that the run of $Q_{39}$ as a function of metal
abundance is not linear in the high metallicity regime.
The fitting second-order polynomial is

\begin{eqnarray}
{\rm [Fe/H]_{UVES}} & = - 2.023 (\pm 0.038) + 5.699\,(\pm 0.589) Q_{39} 
\nonumber \\
    & -3.188\,(\pm 1.320)Q_{39}^2 \nonumber    
\end{eqnarray}
with $\sigma=0.14$ dex and correlation coefficient $r=0.99$ from 14 GCs.

\begin{figure}
\centering
\includegraphics[scale=0.45]{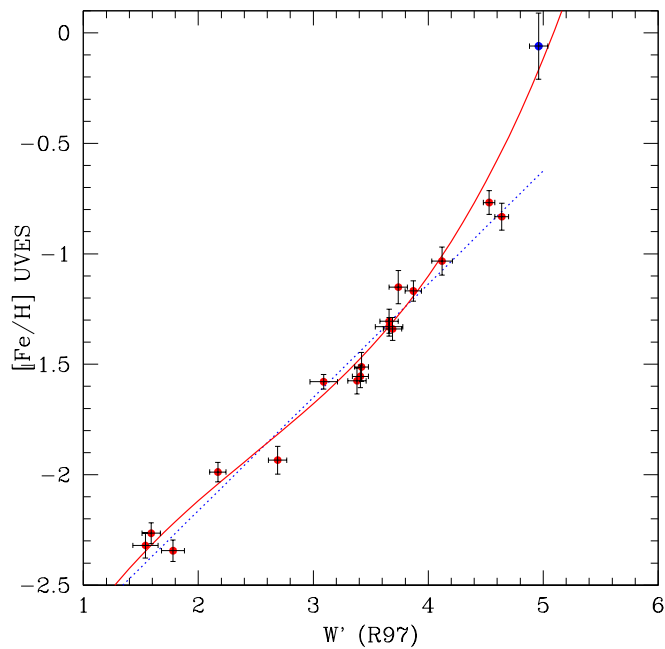}
\caption{Calibration of the reduced strength W' of the calcium triplet from
Rutledge et al. (1997) after adding the metal-rich bulge cluster NGC~6553
(NGC~6528 does not have W' measured by Rutledge et al.) from
Carretta et al. (2001; blue filled square). The blue dotted line is the linear
regression of Fig.~\ref{f:r97u} and the red solid line the cubic spline fit to
the data.}
\label{f:w97}
\end{figure}

Although anchored to only one cluster, NGC~6553, in the near solar metallicity region, the new
calibration of the calcium triplet reduced strength, of the form

\begin{eqnarray}
{\rm [Fe/H]_{UVES}} & = - 3.632 (\pm 0.026) + 1.233\,(\pm 1.014) W' 
\nonumber \\
    & -0.326\,(\pm 0.329)W'^2 +0.044\,(\pm 0.033)W'^3 \nonumber    
\end{eqnarray}
(with $\sigma=0.11$ dex and a correlation coefficient $r=0.99$ from 18 GCS) is a better fit (see
Fig.~\ref{f:w97}) than the simple linear regression previously shown (and indicated by
the dotted line in this Figure).

Moreover, from Fig.~\ref{f:q39} and Fig.~\ref{f:w97} we note that simple
linear extrapolations
would result in [Fe/H] values for clusters with [Fe/H]$>-0.5$ that would be  overestimated from 
the $Q_{39}$ index and underestimated when using the W' index. This fact underlines the risks
of a too crude extrapolation procedure, and we believe that this justifies the inclusion of
NGC~6553 and NGC~6528 among our calibrating clusters.

We used these transformations to derive [Fe/H] values on our scale and the
previous relations of
Sect. 5 to transform metallicities from Carretta and Gratton (1997) and from Kraft and Ivans
(2003) to the present scale.
Finally, we computed the weighted average of the metallicity for each cluster, assuming as 
weights the $rms$ scatters of the fitting relations in each case. For our UVES based metal
abundances we adopted internal errors (see Carretta et al. 2009b) as weighting factor.
 
The resulting metallicity are listed in Tab.~\ref{t:compilation} for 95 GCs in the Harris (1996)
catalogue for which this average was possible. Associated errors are 1$\sigma$ $rms$ of the
weighted average and they represent the errors associated to the metallicity with respect to the
metallicity scale/ranking defined in this paper.

The other 38 entries in the Harris catalogue have no determinations of [Fe/H] from homogeneous
high resolution spectroscopy or compilation of metallicity-dependent
indices. In general, these
are distant clusters, not included in any of the previous main studies. Their metal abundance in
the Harris catalogue is based on a variety of metallicity indicators.
For these objects we proceeded as
follow. The comparison of our weighted average [Fe/H] values with metallicity entries in the
Harris catalogue for 94 clusters in common\footnote{Ter 7 is the most discrepant cluster in
this comparison and is omitted from the fit.} provides a slope equal to unity (when metallicity
from Harris is used as independent variable) with an offset of 0.025 dex. Hence, for the 38
remaining clusters we simply added this offset to the [Fe/H] values listed in the Harris
catalogue (flag 2 in Tab.~\ref{t:compilation}). In this case, the attached error is simply the 
$rms$ scatter from the comparison of the 94 clusters, i.e. 0.09 dex.

\begin{table*}
\caption{Compilation of cluster metallicities on the present scale}
\scriptsize
\centering
\begin{tabular}{lllll c lllll}
\hline
GC      &other&[Fe/H]  & err  &note& ~~~~~&  GC	&other&[Fe/H]  & err  &note\\
\hline
NGC104  &47Tuc&$-$0.76 & 0.02 &  1 & &  Ter2   &      &$-$0.29 & 0.09 &  2 \\
NGC288  &     &$-$1.32 & 0.02 &  1 & &  NGC6366&      &$-$0.59 & 0.08 &  1 \\
NGC362  &     &$-$1.30 & 0.04 &  1 & &  NGC6362&      &$-$1.07 & 0.05 &  1 \\
NGC1261 &     &$-$1.27 & 0.08 &  1 & &  Ter4   &      &$-$1.62 & 0.09 &  2 \\
Pal1	&     &$-$0.51 & 0.09 &  2 & &  HP1    &      &$-$1.57 & 0.09 &  2 \\
AM1	&     &$-$1.84 & 0.09 &  2 & &  LILLER1&      &  +0.40 & 0.09 &  2 \\
Eridanus&     &$-$1.44 & 0.08 &  1 & &  NGC6380& TON1 &$-$0.40 & 0.09 &  2 \\
Pal2	&     &$-$1.29 & 0.09 &  2 & &  Ter1   &      &$-$1.29 & 0.09 &  2 \\
NGC1851 &     &$-$1.18 & 0.08 &  1 & &  NGC6388&      &$-$0.45 & 0.04 &  1 \\
NGC1904 &M79  &$-$1.58 & 0.02 &  1 & &  NGC6402& M14  &$-$1.39 & 0.09 &  2 \\
NGC2298 &     &$-$1.96 & 0.04 &  1 & &  NGC6401&      &$-$1.01 & 0.14 &  1 \\
NGC2419 &     &$-$2.20 & 0.09 &  2 & &  NGC6397&      &$-$1.99 & 0.02 &  1 \\
NGC2808 &     &$-$1.18 & 0.04 &  1 & &  Pal6   &      &$-$1.06 & 0.09 &  2 \\
E3	&     &$-$0.73 & 0.09 &  2 & &  NGC6424&      &$-$2.36 & 0.09 &  2 \\
Pal3	&     &$-$1.67 & 0.08 &  1 & &  Ter5   &      &  +0.16 & 0.09 &  2 \\
NGC3201 &     &$-$1.51 & 0.02 &  1 & &  NGC6440&      &$-$0.20 & 0.14 &  1 \\
Pal4	&     &$-$1.46 & 0.08 &  1 & &  NGC6441&      &$-$0.44 & 0.07 &  1 \\
NGC4147 &     &$-$1.78 & 0.08 &  1 & &  NGC6453&      &$-$1.48 & 0.14 &  1 \\
NGC4372 &     &$-$2.19 & 0.08 &  1 & &  Ter6   &      &$-$0.40 & 0.09 &  2 \\
Rup106  &     &$-$1.78 & 0.08 &  1 & &  UKS1   &      &$-$0.40 & 0.09 &  2 \\
NGC4590 &M68  &$-$2.27 & 0.04 &  1 & &  NGC6496&      &$-$0.46 & 0.07 &  1 \\
NGC4833 &     &$-$1.89 & 0.05 &  1 & &  Ter9   &      &$-$2.07 & 0.09 &  2 \\
NGC5024 &M53  &$-$2.06 & 0.09 &  2 & &  NGC6517&      &$-$1.24 & 0.14 &  1 \\
NGC5053 &     &$-$2.30 & 0.08 &  1 & &  NGC6522&      &$-$1.45 & 0.08 &  1 \\
NGC5139 &OCEN &$-$1.64 & 0.09 &  2 & &  NGC6535&      &$-$1.79 & 0.07 &  1 \\
NGC5272 &M3   &$-$1.50 & 0.05 &  1 & &  NGC6528&      &  +0.07 & 0.08 &  1 \\
NGC5286 &     &$-$1.70 & 0.07 &  1 & &  NGC6539&      &$-$0.53 & 0.14 &  1 \\
AM4	&     &$-$2.07 & 0.09 &  2 & &  NGC6544&      &$-$1.47 & 0.07 &  1 \\
NGC5466 &     &$-$2.31 & 0.09 &  2 & &  NGC6541&      &$-$1.82 & 0.08 &  1 \\
NGC5634 &     &$-$1.93 & 0.09 &  2 & &  NGC6553&      &$-$0.16 & 0.06 &  1 \\
NGC5694 &     &$-$2.02 & 0.07 &  1 & &  NGC6558&      &$-$1.37 & 0.14 &  1 \\
IC4499  &     &$-$1.62 & 0.09 &  2 & &  IC1276 & Pal7 &$-$0.65 & 0.09 &  2 \\
NGC5824 &     &$-$1.94 & 0.14 &  1 & &  NGC6569&      &$-$0.72 & 0.14 &  1 \\
Pal5	&     &$-$1.41 & 0.09 &  2 & &  NGC6584&      &$-$1.50 & 0.09 &  2 \\
NGC5897 &     &$-$1.90 & 0.06 &  1 & &  NGC6624&      &$-$0.42 & 0.07 &  1 \\
NGC5904 &M5   &$-$1.33 & 0.02 &  1 & &  NGC6626&      &$-$1.46 & 0.09 &  2 \\
NGC5927 &     &$-$0.29 & 0.07 &  1 & &  NGC6638&      &$-$0.99 & 0.07 &  1 \\
NGC5946 &     &$-$1.29 & 0.14 &  1 & &  NGC6637& M69  &$-$0.59 & 0.07 &  1 \\
NGC5986 &     &$-$1.63 & 0.08 &  1 & &  NGC6642&      &$-$1.19 & 0.14 &  1 \\
Pal14	&     &$-$1.63 & 0.08 &  1 & &  NGC6652&      &$-$0.76 & 0.14 &  1 \\
NGC6093 &M80  &$-$1.75 & 0.08 &  1 & &  NGC6656& M22  &$-$1.70 & 0.08 &  1 \\
NGC6101 &     &$-$1.98 & 0.07 &  1 & &  Pal8   &      &$-$0.37 & 0.14 &  1 \\
NGC6121 &M4   &$-$1.18 & 0.02 &  1 & &  NGC6681& M70  &$-$1.62 & 0.08 &  1 \\
NGC6144 &     &$-$1.82 & 0.05 &  1 & &  NGC6712&      &$-$1.02 & 0.07 &  1 \\
NGC6139 &     &$-$1.71 & 0.09 &  2 & &  NGC6715& M54  &$-$1.44 & 0.07 &  1 \\
NGC6171 &M107 &$-$1.03 & 0.02 &  1 & &  NGC6717& Pal9 &$-$1.26 & 0.07 &  1 \\
NGC6205 &M13  &$-$1.58 & 0.04 &  1 & &  NGC6723&      &$-$1.10 & 0.07 &  1 \\
NGC6218 &M12  &$-$1.33 & 0.02 &  1 & &  NGC6749&      &$-$1.62 & 0.09 &  2 \\
NGC6229 &     &$-$1.43 & 0.09 &  2 & &  NGC6752&      &$-$1.55 & 0.01 &  1 \\
NGC6235 &     &$-$1.38 & 0.07 &  1 & &  NGC6760&      &$-$0.40 & 0.14 &  1 \\
NGC6254 &M10  &$-$1.57 & 0.02 &  1 & &  NGC6779& M56  &$-$2.00 & 0.09 &  2 \\
NGC6256 &     &$-$0.62 & 0.09 &  2 & &  ARP2   &      &$-$1.74 & 0.08 &  1 \\
Pal15	&     &$-$2.10 & 0.08 &  1 & &  NGC6809& M55  &$-$1.93 & 0.02 &  1 \\
NGC6266 &M62  &$-$1.18 & 0.07 &  1 & &  Pal11  &      &$-$0.45 & 0.08 &  1 \\
NGC6273 &M19  &$-$1.76 & 0.07 &  1 & &  NGC6838& M71  &$-$0.82 & 0.02 &  1 \\
NGC6284 &     &$-$1.31 & 0.09 &  2 & &  NGC6864& M75  &$-$1.29 & 0.14 &  1 \\
NGC6287 &     &$-$2.12 & 0.09 &  2 & &  NGC6934&      &$-$1.56 & 0.09 &  2 \\
NGC6293 &     &$-$2.01 & 0.14 &  1 & &  NGC6981& M72  &$-$1.48 & 0.07 &  1 \\
NGC6304 &     &$-$0.37 & 0.07 &  1 & &  NGC7006&      &$-$1.46 & 0.06 &  1 \\
NGC6316 &     &$-$0.36 & 0.14 &  1 & &  NGC7078& M15  &$-$2.33 & 0.02 &  1 \\
NGC6325 &     &$-$1.37 & 0.14 &  1 & &  NGC7089& M2   &$-$1.66 & 0.07 &  1 \\
NGC6341 &M92  &$-$2.35 & 0.05 &  1 & &  NGC7099& M30  &$-$2.33 & 0.02 &  1 \\
NGC6333 &M9   &$-$1.79 & 0.09 &  2 & &  Pal12  &      &$-$0.81 & 0.08 &  1 \\
NGC6342 &     &$-$0.49 & 0.14 &  1 & &  Pal13  &      &$-$1.78 & 0.09 &  2 \\
NGC6356 &     &$-$0.35 & 0.14 &  1 & &  NGC7492&      &$-$1.69 & 0.08 &  1 \\
NGC6355 &     &$-$1.33 & 0.14 &  1 & &  Ter7   &      &$-$0.12 & 0.08 &  1 \\
NGC6352 &     &$-$0.62 & 0.05 &  1 & &         &      &        &      &    \\		
\hline
\end{tabular}
\label{t:compilation}
\begin{list}{}{}
\item[1:] [Fe/H] values are the weighted average of metallicities from the present
work (UVES values), GC97, KI03, and the recalibration of the Q39 and W' indices
\item[2:] [Fe/H] values are those from the Harris (1996) updated catalogue with
an offset of 0.025 dex added.
\end{list}
\end{table*}

\end{appendix}

\begin{thebibliography}{}

\bibitem[]{} Alonso, A., Arribas, S. \& Martinez-Roger, C. 1999, A\&AS, 140, 261
\bibitem[]{} Alonso, A., Arribas, S. \& Martinez-Roger, C. 2001. A\&A, 376, 1039
\bibitem[]{} Bedin, L., Piotto, G., Anderson, J., Cassisi, S., King, I.R.,
 Momany, Y., Carraro, G. 2004, ApJ, 605, L125 
\bibitem[]{} Bellazzini, M. et al. 2008, AJ, 136, 1147 
\bibitem[]{} Bragaglia, A., Carretta, E., Gratton, R.G. et al. 2001, AJ, 121, 327
\bibitem[]{} Bragaglia, A., et al. 2009, in preparation
\bibitem[]{} Carney, B.W., Latham, D.W., Stefanik, R.P., Laird, J.B. 2008, AJ,
 135, 196
\bibitem[]{} Carretta, E., Gratton R.G. 1997, A\&A Suppl., 121, 95
\bibitem[]{} Carretta, E., Bragaglia, A., Gratton R.G., Leone, F., 
  Recio Blanco, A., Lucatello, S. 2006, A\&A, 450, 523 (Paper I)
\bibitem[]{} Carretta, E., Bragaglia, A., Gratton, R.G., Lucatello, S. 2009b, 
 A\&A, in press (Paper VIII), arXiv:0909.2941
\bibitem[]{} Carretta, E., Bragaglia, A., Gratton, R.G., Lucatello,
 S. Momany, Y. 2007a, A\&A, 464, 927 (Paper II)
\bibitem[]{} Carretta, E., Bragaglia, A., Gratton, R.G., Recio-Blanco, A., 
  Lucatello, S., D'Orazi, V., Cassisi, S. 2009c, in preparation 
\bibitem[]{} Carretta, E., Cohen, J.G., Gratton, R.G., Behr, B.B. 2001, AJ, 122,
 1469 
\bibitem[]{} Carretta, E., Gratton R.G., Bragaglia, A., Bonifacio, P. \& 
  Pasquini, L. 2004b, A\&A, 416, 925 
\bibitem[]{} Carretta, E. et al. 2007b,  A\&A, 464, 939 (Paper IV)
\bibitem[]{} Carretta, E. et al. 2007c, A\&A, 464, 967 (Paper VI)
\bibitem[]{} Carretta, E. et al. 2009a, A\&A, in press (Paper VII), arXiv:0909.2938
\bibitem[]{} Casetti-Dinescu, D.I., Girard, T.M., Herrera, D. et al. 2007, AJ,
 134, 195
\bibitem[]{} Charbonnel, C. 1994, A\&A, 282, 811 
\bibitem[]{} Charbonnel, C. 1995, ApJ, 453, L41
\bibitem[]{} Charbonnel, C., Zahn, J.-P. 2007, A\&A, 467, L15
\bibitem[]{} Cohen, J.G. 1983, ApJ, 270, 654
\bibitem[]{} Cohen, J.G., Frogel, J.A., Persson, S.E. 1978, ApJ, 222, 165
\bibitem[]{} Cummings, I.N. 1999, Journal of Astronomical Data, 5, 2
\bibitem[]{} De Angeli, F. et al. 2005, AJ, 122, 3171
\bibitem[]{} Decressin, T., Baumgardt, H., Kroupa, P., Meynet, G., Charbonnel,
 C. 2008, IAU Symp. 258, D. Soderblom et al. eds, arXiv:0812.2912
\bibitem[]{} D'Ercole, A., Vesperini, E., D'Antona, F., McMillan, S.L.W.,
 Recchi, S. 2008, MNRAS, 391, 825 
\bibitem[]{} Denisenkov, P.A., Denisenkova, S.N. 1989, A.Tsir., 1538, 11
\bibitem[]{} Dinescu, D.I., Girard, T.M., van Altena, W.F. 1999, AJ, 117, 1792
\bibitem[]{} Djorgovski, S., Piotto, G., Capaccioli, M. 1993, AJ, 105, 2148
\bibitem[]{} Eggleton, P.P., Dearborn, D.S.P, Lattanzio, J.C. 2007, IAUS, 239,
 286 
\bibitem[]{} Frogel, J.A., Cohen, J.G., Persson, S.E. 1983, ApJ, 275, 773 
\bibitem[]{} Gnedin, O.Y., Zhao, H.S., Pringle, J.E., Fall, S. M., Livio, M.,
 Meylan, G. 2002, ApJ, 568, L23
\bibitem[]{} Gratton, R.G. 1988, Rome Obs. Preprint Ser., 29
\bibitem[]{} Gratton, R.G., Carretta, E., Claudi, R., Lucatello, S.,
  Barbieri, M. 2003, A\&A, 404, 187 
\bibitem[]{} Gratton, R.G., Lucatello, S., Bragaglia, A., Carretta, E., Momany,
Y., Pancino, E., Valenti, E. 2006,  A\&A, 455, 271 (Paper III) 
\bibitem[]{} Gratton, R.G., Sneden, C., Carretta, E. 2004, ARA\&A, 42, 385
\bibitem[]{} Gratton, R.G. et al. 2007, A\&A, 464, 953 (Paper V) 
\bibitem[]{} Gratton, R.G., Carretta, E., Bragaglia, A., Lucatello, S. 2009, 
  A\&A, submitted
\bibitem[]{} Gratton, R.G. et al. 2001, A\&A, 369, 87
\bibitem[]{} Gray, D.F., Carney, B.W., Yong, D. 2008, AJ, 135, 2033
\bibitem[]{} Harris, W.~E. 1996, AJ, 112, 1487
\bibitem[]{} Ita, Y., et al. 2002, MNRAS, 337, L31
\bibitem[]{} Kraft, R.P., Ivans, I.I. 2003, PASP, 115, 143 
\bibitem[]{} Kurucz, R.L. 1993 CD-ROM 13, Smithsonian Astrophysical 
 Observatory, Cambridge
\bibitem[]{} Langer, G.E., Hoffman, R., \& Sneden, C. 1993, PASP, 105, 301
\bibitem[]{} Larson, R.B. 1987 in ``Nearby normal galaxies", ed. S. Faber (NY:
 Springer), p. 26
\bibitem[]{} Marino, A.F., Milone, A., Piotto, G., Villanovo, S., Bedin, L.,
 Bellini, A., Renzini, A. 2009, arXiv0905.4058 
\bibitem[]{} Melendez, J., Cohen, J.G. 2009, ApJ, 699, 2017 
\bibitem[]{} Minniti, D., Geisler, D., Peterson, R.C., Claria, J.J. 1993, ApJ,
 413, 548 
\bibitem[]{} Piotto, G. et al. 2005, ApJ, 621, 777 
\bibitem[]{} Prantzos, N., Charbonnel, C. 2006, A\&A, 458, 135
\bibitem[]{} Ramirez, S.V., Cohen, J.G. 2002, AJ, 123, 3277 
\bibitem[]{} Recio Blanco, A., Aparicio, A., Piotto, G., De Angeli, F., 
 Djorgovski, S.G. 2006, A\&A, 452, 875
\bibitem[]{} Rosenberg A., Saviane I., Piotto G., Aparicio A., 1999, AJ, 118, 2306 
\bibitem[]{} Rutledge, G.A., Hesser, J.E., Stetson, P.B. 1997, PASP, 109, 907
\bibitem[]{} Schwarzschild, M. 1975, ApJ, 195, 137
\bibitem[]{} Searle, L., Zinn, R. 1978, ApJ, 225, 357
\bibitem[]{} Smith, G.H., Martell, S. 2003, PASP, 115, 1211
\bibitem[]{} Smith, V.V., Cunha, K., Ivans, I.I., Lattanzio, J.C.,
 Campbell, S., Hinkle, K.H. 2005, ApJ, 633, 392
\bibitem[]{} Soszynski, I., Udalski, A., Kubiak, M., Szymanski, M., 
 Pietrzynski, G., Zebrun, K., Szewczyk, O., Wyrzykowski, L.2004, Acta
 Astronomica, 54, 129
\bibitem[]{} Suntzeff, N.B., Kraft, R.P. 1996, AJ, 111, 1913
\bibitem[]{} Ventura, P. D'Antona, F., Mazzitelli, I., Gratton, R. 2001, ApJ, 
 550, L65
\bibitem[]{} Wheeler, J.C., Sneden, C., Truran, J.W., Jr. 1989, ARA\&A, 27, 279
\bibitem[]{} Yong, D., Grundahl, F., Johnson, J.A., Asplund, M. 2008a, ApJ, 684,
 1159 
\bibitem[]{} Yong, D., Melendez, J., Cunha, K., Karakas, A.I., Norris,
 J.E., Smith, V.V. 2008b, ApJ, 689, 1020 
\bibitem[]{} Zinn, R, West, M.J. 1984, ApJS, 55 45

\end{thebibliography}
\end{document}